\documentclass[journal]{IEEEtran}
\newcommand{\RNum}[1]{\uppercase\expandafter{\romannumeral #1\relax}}
\usepackage{mathrsfs}
\usepackage{amssymb}
\usepackage[mathcal]{euscript}
\usepackage{diagbox}
\usepackage{cite}
\usepackage[T1]{fontenc}
\usepackage{graphicx}
\usepackage{makecell}
\usepackage{psfrag}
\usepackage{url}
\usepackage{stfloats}
\usepackage{amsmath}
\usepackage{array}
\usepackage{float}
\usepackage{multirow} 
\usepackage{amsthm}
\usepackage{color}
\usepackage{multicol}
\usepackage{stfloats}
\usepackage{enumerate}
\usepackage{subfigure}
\usepackage{booktabs}  

\allowdisplaybreaks[4]
\usepackage[ruled]{algorithm2e}
\usepackage{algorithmic} 
\ifCLASSINFOpdf
\else
\fi
\begin{document}
%
\title{Interactive AI with Retrieval-Augmented Generation for Next Generation Networking}



%
	\author{Ruichen Zhang*,  Hongyang Du*, Yinqiu Liu*, Dusit Niyato,~\IEEEmembership{Fellow,~IEEE}, Jiawen Kang, Sumei~Sun,~\IEEEmembership{Fellow,~IEEE}, Xuemin (Sherman) Shen,~\IEEEmembership{Fellow,~IEEE}, and H. Vincent Poor,~\IEEEmembership{Life Fellow,~IEEE}

\thanks{R. Zhang, H. Du, Y. Liu, and D. Niyato are with the School of Computer Science and Engineering, Nanyang Technological University, Singapore (e-mail: ruichen.zhang@ntu.edu.sg, hongyang001@e.ntu.edu.sg, yinqiu001@e.ntu.edu.sg, dniyato@ntu.edu.sg).}

\thanks{J. Kang is with the School of Automation, Guangdong University of Technology, China (e-mail: kavinkang@gdut.edu.cn).}

\thanks{S. Sun is with  the Institute for Infocomm Research, Agency for Science, Technology and Research, Singapore (e-mail: sunsm@i2r.a-star.edu.sg).}

\thanks{X. Shen is with the Department of Electrical and Computer Engineering, University of Waterloo, Canada (e-mail: sshen@uwaterloo.ca).}

\thanks{H. V. Poor is with the Department of Electrical and Computer Engineering, Princeton University, Princeton, NJ08544, USA (e-mail: poor@princeton.edu).}

\thanks{* means equal contribution.}
}
\maketitle

\begin{abstract}
With the advance of artificial intelligence (AI), the emergence of Google Gemini and OpenAI Q* marks the direction towards artificial general intelligence (AGI). To implement AGI, the concept of interactive AI (IAI) has been introduced, which can interactively understand and respond not only to human user input but also to dynamic system and network conditions. In this article, we explore an integration and enhancement of IAI in networking. We first comprehensively review recent developments and future perspectives of AI and then introduce the technology and components of IAI. We then explore the integration of IAI into the next-generation networks, focusing on how implicit and explicit interactions can enhance network functionality, improve user experience, and promote efficient network management. Subsequently, we propose an IAI-enabled network management and optimization framework, which consists of environment, perception, action, and brain units. We also design the pluggable large language model (LLM) module and retrieval augmented generation (RAG) module to build the knowledge base and contextual memory for decision-making in the brain unit. We demonstrate the effectiveness of the framework through case studies. Finally, we discuss potential research directions for IAI-based networks.
\end{abstract}

\section{Introduction}

With the current development trajectory, artificial intelligence (AI) is moving towards an important stage of realizing artificial general intelligence (AGI). The evolution of AI from its original rule-based algorithms to the adoption of advanced learning models marks a significant shift in the field of computing technology. This process is driven by the increasing growth and complexity of data, requiring more sophisticated AI solutions. While early AI models provided the foundation for modern data-driven environments, their limitations in handling dynamic and large-scale data have prompted the development of more advanced and novel methods. Among these advancements, such as Google Gemini\footnote{https://deepmind.google/technologies/gemini/\#introduction} and OpenAI Q*\footnote{https://community.openai.com/t/what-is-q-and-when-we-will-hear-more/521343}, human-in-the-loop (HITL) systems highlight the importance of integrating human insights into AI decision-making \cite{xu2023transitioning}. These systems combine human feedback with the AI’s learning process, improving the accuracy and contextual understanding of AI results. For example, a 2018 Stanford study showed that AI combined with HITL outperformed human or AI analysis alone in healthcare\footnote{https://scopeblog.stanford.edu/2018/05/08/artificial-intelligence-in-medicine-predicting-patient-outcomes-and-beyond/}. However, HITL systems have limitations in adaptability and real-time responsiveness because they rely on human input, which can be limited, unpredictable, and erroneous.

To address these challenges and effectively develop AGI,  interactive AI (IAI) has been proposed. The primary difference between IAI and HITL lies in their interaction modes, i.e.,  IAI emphasizes immediate and direct interaction between AI and users, whereas  HITL focuses on human participation and supervision in the AI decision-making process \cite{xu2023transitioning}.  IAI systems are capable of instantaneously understanding user inputs, such as voice commands, text messages, or other interactive commands, and intelligently responding or executing tasks based on these inputs. This ability not only enhances user experience but also increases flexibility and effectiveness of AI applications in dynamic environments.  Integrating  IAI with technologies such as retrieval-augmented generation (RAG) and LangChain can further personalize network operations \cite{lewis2020retrieval}. For example, RAG enables IAI systems to extract information from vast databases to enrich their responses and decisions. Combined with LangChain, which extends AI reasoning capabilities, IAI can provide more context-aware solutions based on the existing databases. The main advantages of IAI over the limitations of HITL systems, especially when augmented with RAG and LangChain, include:
\begin{itemize}

\item {\bf Customizabilty and Personalizability:}   IAI, enhanced by RAG and LangChain, offers tailored solutions by aligning AI responses closely with user preferences and needs, resulting in more personalized and user-centric AI applications.
\item {\bf Better Flexibility:} IAI's direct user interaction, supported by LangChain's extended reasoning, can offer more flexibility  to adapt to different network scenarios and users' requirements, enhancing the system’s versatility and satisfaction.

\item {\bf Less  Bias:} The combination of IAI with RAG and LangChain minimizes the reliance on human intervention, thus reducing the potential for human bias in AI decisions and leading to more objective outcomes.
\end{itemize}

It is particularly important to integrate IAI into wireless networks, given its dynamic nature and requirement to continuously adapt the dynamic changes. Fortunately, the capabilities of IAI are particularly promising in addressing these challenges. Its interactive adaptive resource management can optimize the utilization of network resources and improve network performance. For example, in a network with changing user requirements, IAI can dynamically allocate bandwidth to maintain high performance given instantaneous user experience feedback. Although the integration of IAI into networking has several potential advantages, the following issues need to be addressed:
\begin{itemize}
\item {\bf Q1:} Why is IAI suitable for networking?

\item {\bf Q2:} Which networking challenges can IAI address?

\item {\bf Q3:} How can IAI with RAG be applied in networking?
\end{itemize}

Therefore, in this article, we attempt to provide forward-looking research to answer the above ``Why, Which, and How" questions. \textit{To the best of the authors' knowledge, the synergy between IAI and networking is still an open issue.} The contributions of this article are summarized as follows.
\begin{itemize}
\item {\bf A1:} We first review some aspects of the development of AI, then introduce the features, technologies, and composition of IAI, and finally overview the IAI on networking.

\item {\bf A2:} We explore the integration of IAI in networking, focusing on how both implicit and explicit interactions enhance network functionalities, improve user experience, and facilitate efficient network management.

\item {\bf A3:} We construct an IAI-enabled network management and optimization framework. It consists of environment, action, brain, and perception. More importantly, we design pluggable LLM and RAG modules to build the knowledge base and contextual memory for decision-making. Simulation results based on a real network optimization case study verify the effectiveness of the proposed framework.
\end{itemize}

\begin{figure*}[!t]
\centering
\includegraphics[width=\textwidth]{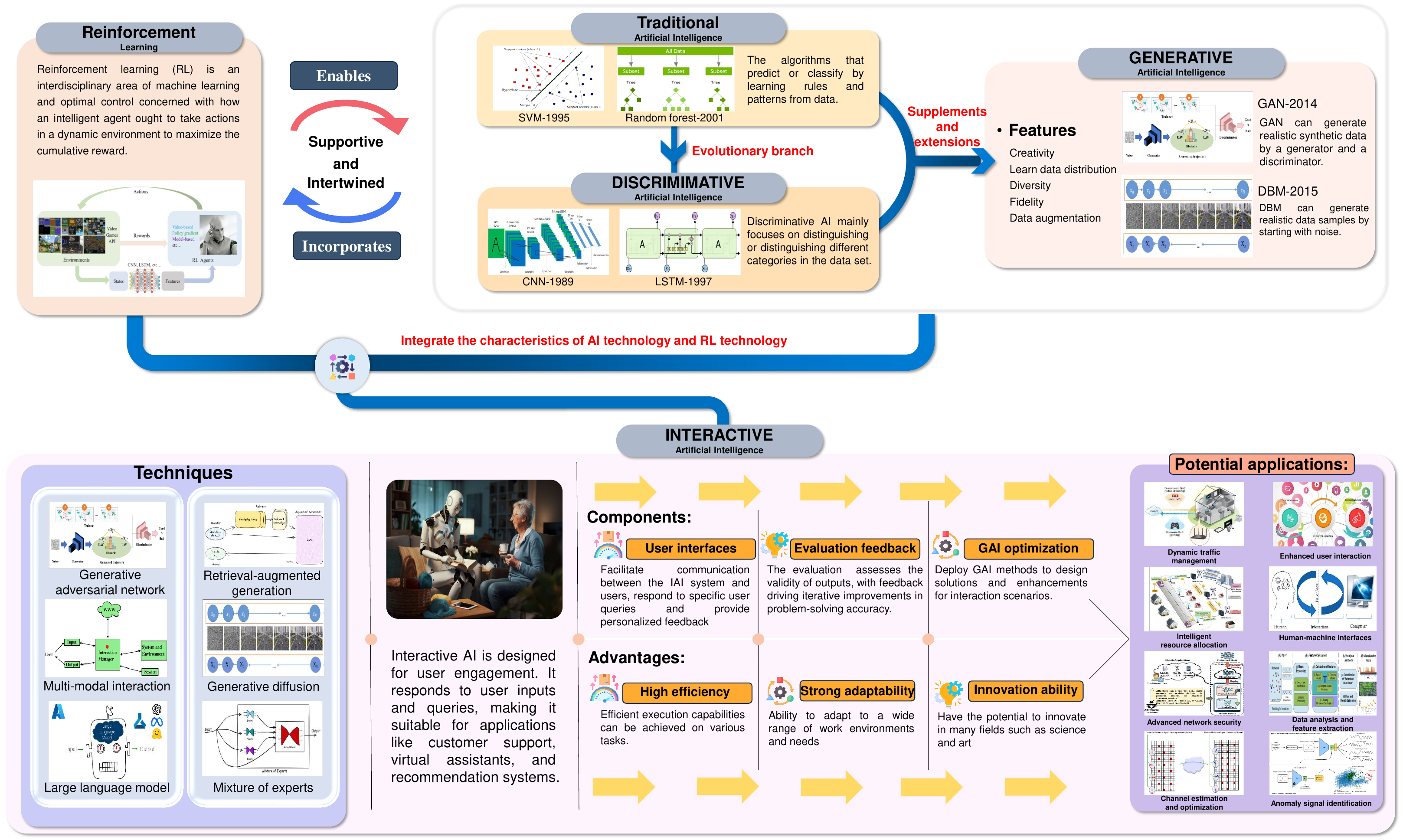}
\caption{The evolution of AI. For IAI, we highlight the role of the mixture of experts, large language models, deep reinforcement learning, retrieval-augmented generation, and generative AI in promoting adaptability and interaction with users. Additionally, the components and advantages of IAI are also presented, emphasizing its efficiency and adaptability in different applications such as optimization and traffic management. }
\label{fig_PSNC_1}
\end{figure*}

\section{Overview of IAI and Networking} 
This section provides an overview of IAI, where some abbreviations are summarized in Table~\ref{table_generative AI_1}.

\subsection{The development of AI}


{\bf Phase 1: Traditional Artificial Intelligence (TAI)} TAI consists of rule-based systems designed for specific tasks within well-defined parameters\cite{lu2019artificial}. It leverages algorithms such as support vector machines (SVMs) and other fundamental methods to excel at pattern recognition and basic prediction tasks. Although TAI systems are effective at processing structured data, their limited flexibility highlights the need for more dynamic AI approaches, leading to the development of discriminative AI/predictive AI.

\begin{table}[!t]
\centering
\caption{SUMMARY OF ABBREVIATIONS.} 
\includegraphics[width= 0.49\textwidth]{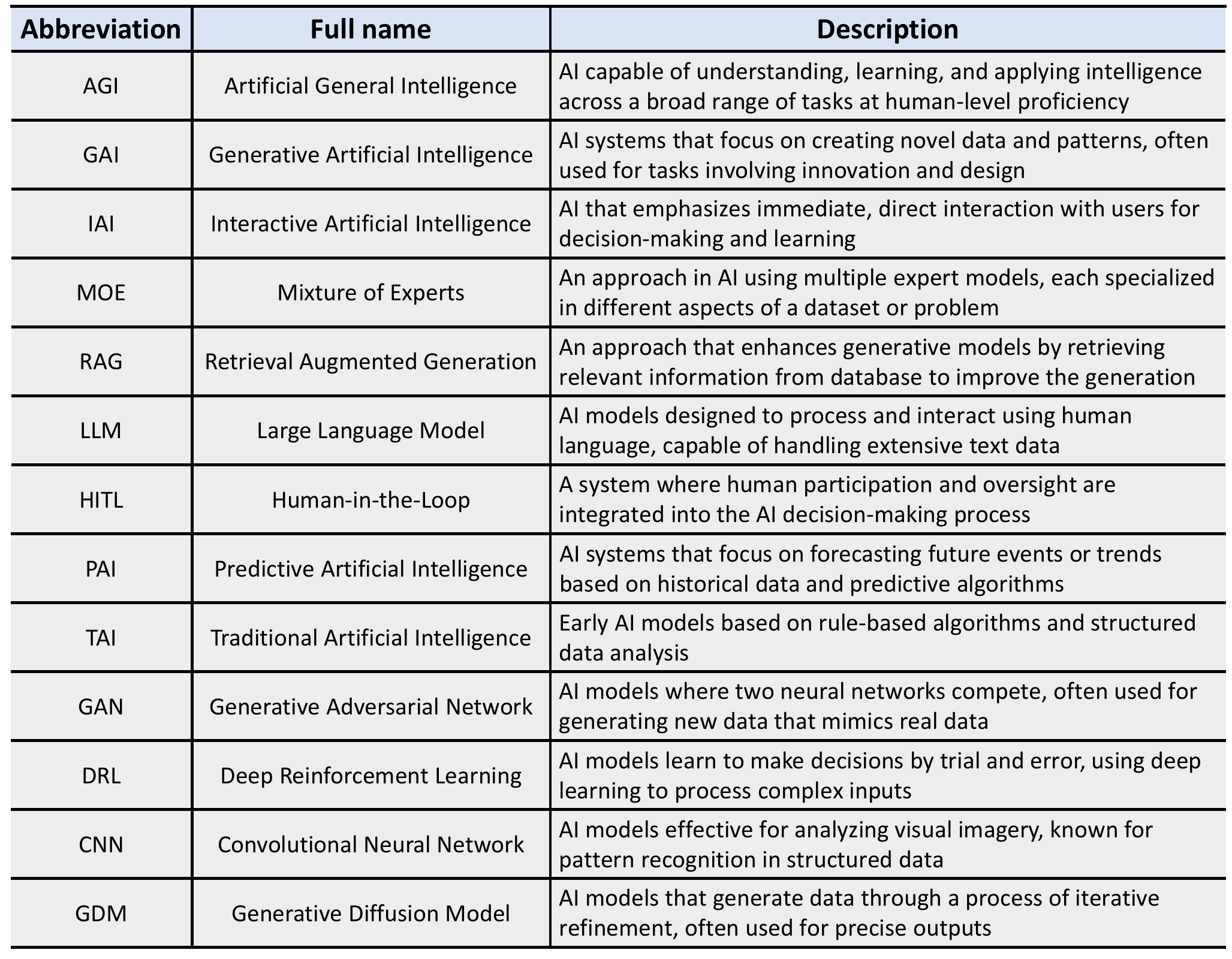}
\label{table_generative AI_1}
\end{table}

{\bf Phase 2: Predictive AI (PAI) /Discriminative AI (DAI)} With the use of deep neural networks such as convolutional neural networks (CNNs), PAI and DAI are good at learning special paradigms from large data sets \cite{li2021survey}. PAI/DAI, with their enhanced learning algorithms, advances beyond the capabilities of TAI. However, their reliance on extensively annotated datasets, primarily for classification and prediction, has practical limitations, leading to the emergence of GAI.

{\bf Phase 3: Generative Artificial Intelligence (GAI)} GAI marks a new era where AI systems can create new data and patterns, showcasing a degree of creativity \cite{du2023beyond}. With the advent of generative diffusion models (GDMs) and generative adversarial networks (GANs), AI's role expands from analysis to creation, which can produce innovative outputs and simulations. However, GAI's reliance on pre-existing data patterns has limitations such as generating inaccurate results, leading to the emergence of IAI.

{\bf Phase 4: Interactive Artificial Intelligence (IAI)} IAI goes beyond traditional data interaction paradigms by engaging human users in dynamic and reciprocal information exchanges. This stage allows the AI system to contextualize interactions and learn from user input, thus facilitating adaptive feedback mechanisms\cite{teso2023leveraging}. As a result, AI systems can bridge the gap between AI and human interaction and then move towards AGI.

For clarity, the evolution of AI development is illustrated in Fig.~\ref{fig_PSNC_1}.

\subsection{Overview of IAI}
IAI is based on the principle of dynamic information exchange to meet the needs of administrators and human end-users, where the foundation of IAI lies in the following AI technologies:

{\bf Retrieval-Augmented Generation (RAG):} 
RAG plays a pivotal role in IAI framework by merging retrieval-based and generative AI techniques. RAG allows IAI systems to access a vast external knowledge base, retrieving relevant information to augment the generation process\cite{lewis2020retrieval}.

{\bf Large Language Models (LLMs):}  LLMs are a cornerstone of IAI, particularly due to their capacity to process and generate human-like text, which facilitates complex interactive dialogues. This ability is central to IAI's focus on enhancing user interaction, as LLMs enable AI systems to comprehend and respond to natural language inputs in a conversational manner, thereby significantly improving the interactive and intuitive nature of AI systems \cite{chen2023octavius}.

{\bf Multi-modal Interaction:} IAI systems are not limited to a single modal; they are adept at understanding and responding to various input types, such as text, voice, and images \cite{zhang2023generative}. This multi-modal interaction capability is fundamental to IAI, embodying the principle of versatility and adaptability in interactions. It ensures that AI systems can engage with users in the most natural and intuitive ways possible, adapting to different interaction modes seamlessly.

{\bf Mixture of Experts (MOE):} The MOE framework is an integral component of IAI that embodies its responsiveness to specialized situation-awareness. MOE models are composed of multiple expert sub-models, each trained to handle different aspects or subsets of the data\cite{chen2023octavius}. This specialization enables IAI systems to leverage the collective expertise of these individual models to solve complex problems. The gating mechanism is the core feature of MOE, which dynamically determines the relevance of each expert to a given input, thereby directing the input to the most appropriate expert.

{\bf Deep Reinforcement Learning (DRL):} DRL embodies the interactivity and adaptability of IAI. It revolves around learning from interactions with the environment and making decisions based on feedback, such as rewards and penalties\cite{zhang2023energy}. This principle aligns well with the idea that IAI learns and develops through continuous interaction with the user or environment.

{\bf Generative Diffusion Models (GDMs):} GDMs improve iteratively through interactions, consistent with the principles of IAI. These models generate data by learning to reverse a diffusion process, which essentially involves a series of interacting steps\cite{du2023beyond}. Each iteration in the process presents an opportunity for the model to adapt and refine its output, emphasizing the IAI principle of continuous evolution and response to new information.

{\bf Generative Adversarial Networks (GANs):} 
GANs, with their dueling network structure, inherently align with IAI's principles of interactive learning. The continuous learning process within GANs, where each network learns from the other, enables the system to adaptively refine its outputs \cite{zhang2023generative}.

\subsection{Features of IAI}

In the networking domain, IAI systems are characterized by three main components, each of which is an integral part of its functionality as follows:

\begin{table*}[!t]
\centering
\caption{Summary of potential issues of IAI.} 
\includegraphics[width=\textwidth]{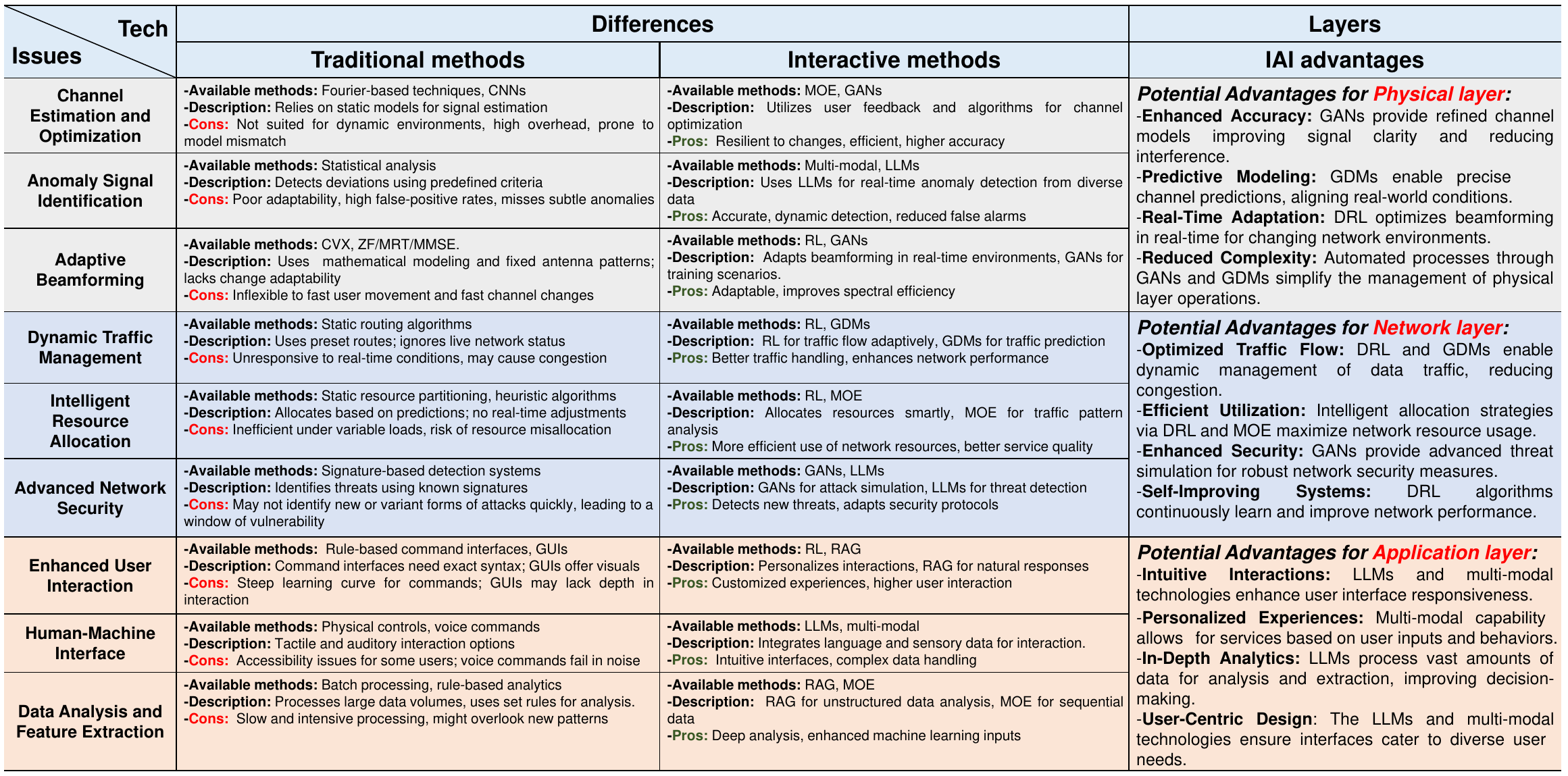}
\label{table_generative AI}
\end{table*}
{\bf User Interfaces:}  Modern IAI systems have revolutionized user interfaces (UIs), extending beyond conventional text and graphical interfaces. These systems incorporate LLMs and multi-modal techniques, allowing users to interact with network management systems through conversational language. A notable example is Agent GPT\footnote{https://agentgpt.reworkd.ai/de}, which employs LLMs and multi-modal capability for interactive user interfaces in a dynamic plan-execute pattern. Unlike standard UIs, Agent GPT engages proactively with users, providing clarity for ambiguous instructions and continually modifying its approach based on user feedback.

{\bf Intelligent Evaluation and Adaptability:} Network environments benefit from IAI systems that not only assess performance using real-time analytics and simulation-based testing but also adapt and evolve over time. These systems integrate cognitive network operations to create a feedback loop, continuously refining network configurations based on user inputs and environmental data. This enables AI models to develop their understanding strategies and enhance their decision-making processes. Google's VirusTotal Code Insight\footnote{https://blog.virustotal.com/2023/04/introducing-virustotal-code-insight.html} is one such system that employs an LLM and MOE for script analysis, thus enabling human users to identify potential threats through responsive feedback during use.

{\bf Advanced Generative AI and Network Optimization:} The core of IAI systems is the deployment of state-of-the-art GAI. These systems utilize algorithms like GDM and TBM for network design optimization. They facilitate the automated generation of network topologies, simulate traffic for congestion forecasting, and provide routing optimization solutions. For instance, Google's B4 network\footnote{https://www.b4networks.ca/} employs a combination of MOE and GAN techniques for capacity planning and traffic engineering, ensuring optimal data flow and resource utilization. The generative process in these systems is iterative, leveraging both historical and real-time data to proactively adapt to evolving network conditions and user behavior trends.

\subsection{IAI for Networking}

IAI significantly benefits networking by integrating intelligence across multiple operational layers. It enhances networks from the physical layer to the application layer.

{\bf Physical Layer:}
In the physical layer, IAI can be used to improve network efficiency and reliability. A typical example is the application of channel estimation and optimization. IAI, particularly DRL approaches, leverage feedback-oriented neural network interaction to iteratively optimize beamforming patterns based on real-time environmental feedback, improving modulation techniques for robust communication \cite{zhang2023energy}.  This leads to more efficient use of the spectrum, reduced interference, and enhanced signal quality, particularly in dynamic environments where channel conditions frequently change.

 {\bf Network Layer:}
In the network layer, IAI can be used to streamline network operations and enhance performance. A typical example is the application of dynamic network traffic management. IAI, particularly DRL and GDM, can dynamically manage and optimize network traffic flows by continuously learning and adapting to traffic conditions \cite{du2023beyond}. This includes predictive traffic modeling, real-time congestion management, and intelligent routing strategies, resulting in reduced latency, optimized bandwidth usage, and improved overall network resilience.

{\bf Application Layer:}
In the application layer, IAI can be used to enhance user interaction and experience. A typical example is the application of human-machine interfaces. IAI, particularly LLMs, can process and generate natural language for user interaction, while multi-modal methods integrate visual, auditory, and other data forms to enrich the interface \cite{xi2023rise}. This leads to more personalized and engaging user interactions, better accessibility for diverse user groups, and an overall more responsive and intelligent application environment.

For clarity,  the potential issues of IAI in different layers are summarized in Table~\ref{table_generative AI}.

\section{Interactive Applications in Networking }

In this section, we explore the integration of IAI in networking, focusing on how both implicit and explicit interactions. For clarity, the summary of implicit and explicit interaction is shown in Table~\ref{table_generative AI_v2}.

\begin{table*}[!t]
\caption{Summary of Implicit and Explicit Interaction in Networking.} 
\includegraphics[width=\textwidth]{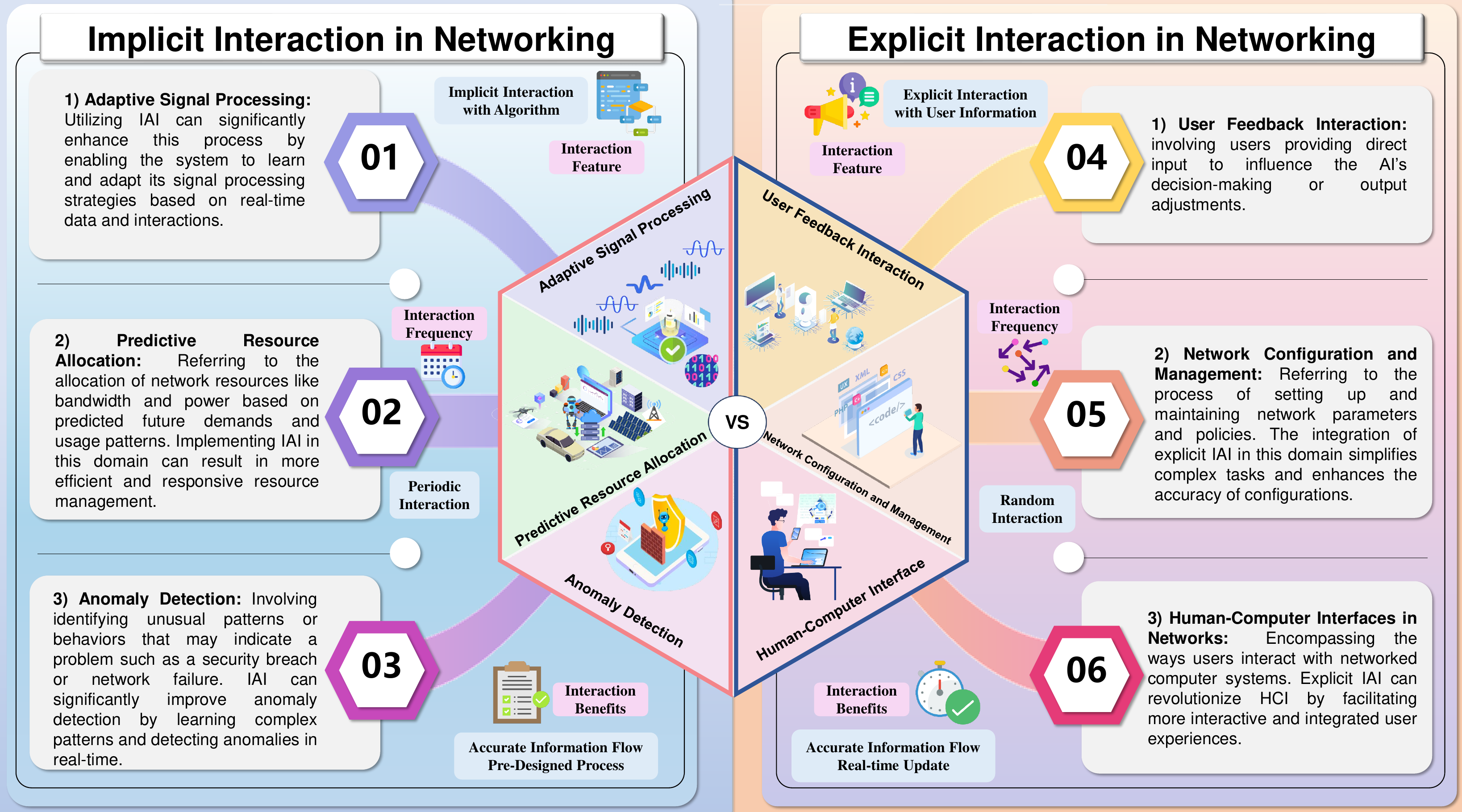}
\label{table_generative AI_v2}
\end{table*}

\subsection{Implicit Interaction in Networking}

Implicit interaction in networking refers to the process where AI systems engage and adapt to their environment, including network conditions and user behaviors, without direct external inputs or commands. The benefits of such interactions include increased system autonomy, improved adaptability to changing conditions, and enhanced overall network efficiency \cite{8017430,zhang2023energy, chen2021learning}. Some major applications of implicit interaction in networking are as follows:

{\bf Adaptive Signal Processing:} Adaptive Signal Processing in wireless communications involves dynamically adjusting signal processing algorithms in response to changing network conditions. Utilizing IAI can significantly enhance this process by enabling the system to learn and adapt its signal-processing strategies based on real-time data and interactions. For example, in \cite{8017430}, the authors proposed an IAI framework for steganographic distortion learning framework. This framework employs a GAN composing two subnetworks, i.e., a steganographic generator and a steganalytic discriminator, which through adversarial training, learns the probabilities of embedding changes in pixels to minimize detectability. In this framework, implicit interactions in the form of adaptive learning allow the model to evolve from simple random embedding to advanced content-adaptive embedding, significantly improving its security performance with each iteration.  Simulation results showed after 180,000 training iterations, the IAI model's system performance steadily improved, achieving an average embedding rate close to the targeted capacity.

{\bf Predictive Resource Allocation:}  Predictive Resource Allocation refers to the allocation of network resources like bandwidth and power based on predicted future demands and usage patterns. Implementing IAI in this domain can result in more efficient and responsive resource management. For example, in \cite{zhang2023energy},  the authors proposed a DRL-based IAI method to optimize system energy efficiency (EE) by adjusting beamforming vectors, power splitting ratios, and phase shifts, considering users' quality of service (QoS) and the transmitter's power constraints. The proposed IAI method implicitly interacted with the whole network environment, learning to fine-tune these parameters without direct human input to satisfy energy harvesting and communication QoS requirements.  Simulation results reveal that this approach yields EE close to an upper bound scheme (i.e., about 2\% performance gap) while significantly reducing computation time (i.e., five orders of magnitude), particularly in dynamic wireless conditions. 

{\bf Anomaly Detection:} Anomaly Detection in networking involves identifying unusual patterns or behaviors that may indicate a problem, such as a security breach or network failure. IAI can significantly improve anomaly detection by learning complex patterns and detecting anomalies in real-time. For example, in \cite{chen2021learning}, the authors proposed an IAI framework to automatically identify dependencies between sensors for anomaly detection in multivariate time series data. The IAI framework employed a  Gumbel-Softmax Sampling-based connection learning policy to automatically learn graph structures depicting sensor dependencies, and integrates this with graph convolutions and an IAI architecture for efficient anomaly detection. The framework's implicit interaction involved determining bi-directional connections among sensors, where high probability values imply strong connections, enhancing the system's ability to predict and identify anomalies.  This method is effective as it incorporates user feedback and network interactions to refine its anomaly detection capability continually. Simulation results prove that the TBM-based IAI framework improves precision, recall, and F1-score by at least 6\%, 16\%, and 11\%, respectively, compared with traditional methods in anomaly detection. 

\textit{Lesson learned:} 
The exploration of implicit IAI in networking underscores the integral role of automated adaptability and self-learning in enhancing network efficiency and security. A major benefit of implementing implicit IAI is its proficiency in navigating the complexities of dynamic network environments. Implicit IAI, through mechanisms like machine learning algorithms and predictive analytics, excels in discerning and adjusting to these changes autonomously.  For instance, in adaptive signal processing, IAI has shown a notable improvement in efficiency, with performance enhancements measurable at approximately 10-15\% in complex scenarios. In resource allocation, IAI has not only achieved near-optimal energy efficiency but also accelerated convergence speed by up to five times compared to traditional methods. Additionally, in anomaly detection, IAI's continuous learning from network interactions has led to a substantial increase in accuracy, outperforming traditional methods by at least 6\% to 16\% in key metrics. These improvements highlight IAI’s real-time adaptability, faster processing, and reduced reliance on large data sets.

\subsection{Explicit  Interaction in Networking}

Explicit interaction in networking involves deliberate and direct engagement between users or network administrators and the AI system. This type of interaction facilitates precise control over network functionalities and enhances user involvement in decision-making processes, leading to more accurate and user-centric outcomes \cite{dhole2023interactive, wang2023making, singh2023hide}. Some major applications of explicit interaction in networking are as follows:

{\bf User Feedback Interaction:} User feedback interaction involves users providing direct input to influence the AI's decision-making or output adjustments. IAI enhances this process by enabling more intuitive and responsive interactions, where user feedback can dynamically shape AI responses or network adjustments. For example,  in \cite{dhole2023interactive}, the authors introduced the Query Generation Assistant, a search interface that uses the LLM-based IAI method to facilitate interactive and automatic query generation in challenging search scenarios such as cross-lingual retrieval. This system allows users to actively interact in refining queries generated by LLMs, providing feedback and edits throughout the process. It significantly improves qualitative analysis in complex search tasks and is a valuable tool for conducting human-in-the-loop simulations.

{\bf Network Configuration and Management:}  Network configuration and management refers to the process of setting up and maintaining network parameters and policies. The integration of explicit IAI in this domain simplifies complex tasks and enhances the accuracy of configurations. For example,  in \cite{wang2023making},  the authors explored the use of the LLM-based IAI method to manage network configuration tasks. They introduce NETBUDDY, a system that translates high-level natural language requirements into low-level network configurations. This method involves explicit interactions where NETBUDDY decomposes the translation process into multiple steps, ensuring more accurate and efficient configurations, and demonstrating considerable improvement in simplifying and automating network management tasks. The simulation results showed that the method adopted not only ensures network accuracy but also increases efficiency by approximately 6 times compared with traditional ones.

{\bf Human-Computer Interfaces:}  Human-computer interfaces (HCI) in networking encompasses the ways users interact with networked computer systems. Explicit IAI can revolutionize HCI by facilitating more interactive and integrated user experiences. For example,  in \cite{singh2023hide},   the authors investigated machine-in-the-loop creative writing using a multi-modal based IAI method that suggested improvements through sight, sound, and language, marking an innovative approach in AI-assisted creativity. This method engaged human users in explicit interactions, where they incorporated AI-generated suggestions into their writing, a process involving adaptability and cognitive integration.  Simulations effectively demonstrated the potential of IAI method to enhance the creativity of human-machine collaboration.

\textit{Lesson learned:} 
Explicit IAI interactions in networking emphasize the critical role of user involvement in shaping AI-driven functionalities. This approach highlights the integration of user feedback into AI systems, resulting in more precise and customized network solutions. Unlike conventional AI, which relies on predefined algorithms, explicit IAI uses user input to refine its decision-making. This direct engagement aligns AI outputs with user-specific needs and preferences. For instance, digital twins and semantic communications can use explicit IAI to interpret and prioritize data based on its meaning and relevance to user requirements. This not only ensures accuracy but also personalizes network operations to individual contexts. Furthermore, explicit IAI is able to create a collaborative environment, merging human expertise with AI capabilities (i.e., MOE), thereby boosting the overall efficiency and effectiveness of network management.

\section{Case Study: IAI-enabled problem formulation framework}

In this section, we propose a framework that utilizes an IAI agent with RAG to help network users and designers formulate optimization problems in the network domain.

\subsection{Motivations}

\begin{figure*}[!t]
\centering
\includegraphics[width=\textwidth]{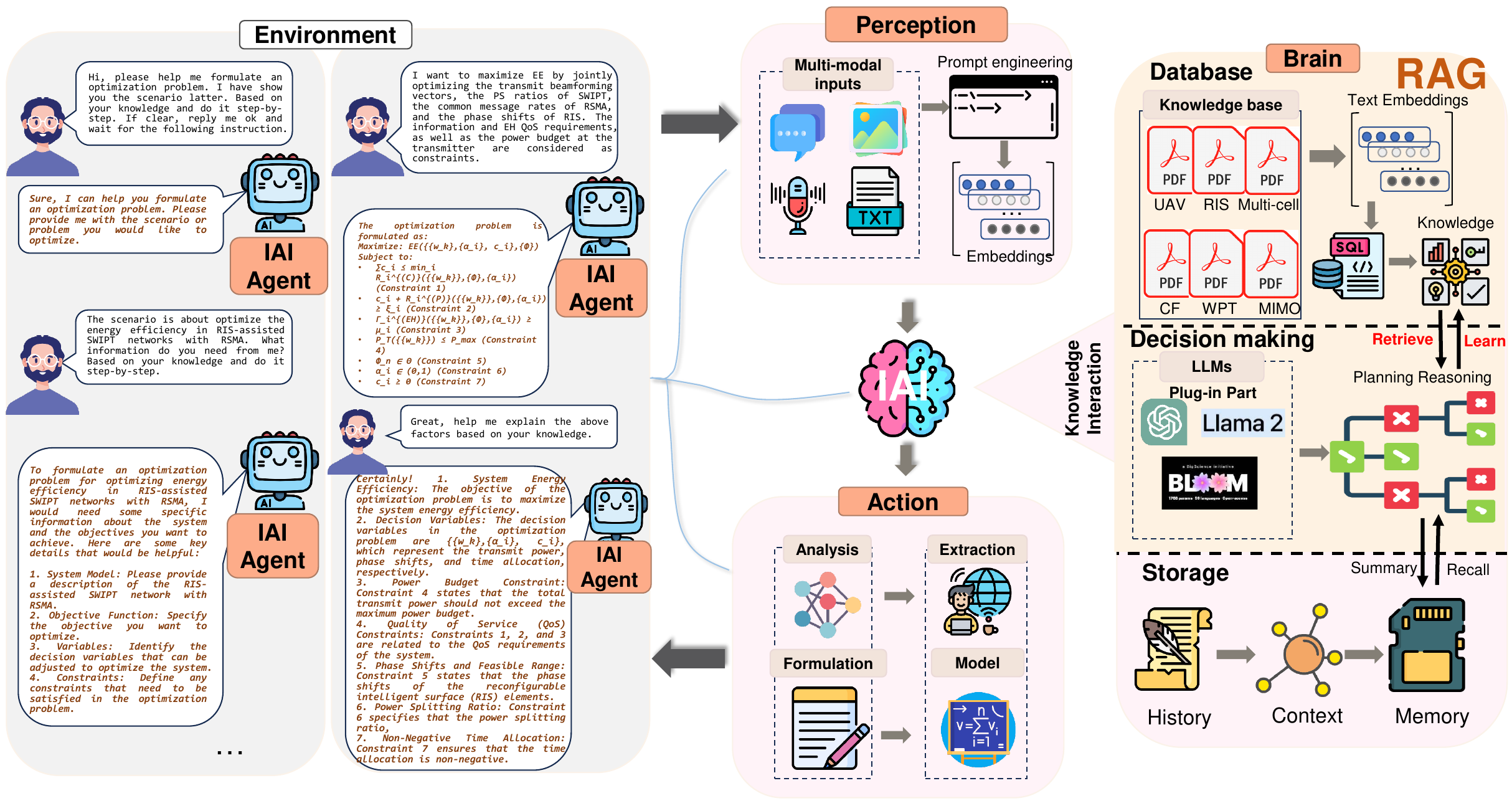}
\caption{In the IAI-enabled problem formulation framework, three key components are presented: brain, perception, and action. The brain units with LLM and RAG modules as the central processor, handling essential tasks such as memory retention, retrieval processing, and making decisions. Meanwhile, the perception module is responsible for acquiring and interpreting diverse data from the environment. Lastly, the action module implements responses and interacts with the environment, utilizing various tools for execution.}
\label{fig_PSNC_2}
\end{figure*}
In wireless network resource allocation, modeling complex real-world scenarios as mathematical optimization problems have traditionally required a deep understanding of complex equations and methods. This task can be challenging, particularly for newcomers or those with interdisciplinary backgrounds.

Fortunately, the IAI framework offers a transformative solution to these challenges. By interpreting the network environment and goals as defined by designers, the IAI system automatically formulates the appropriate optimization problem. This advancement speeds up the problem formulation process and importantly ensures that the optimization models accurately reflect the intricate details of network scenarios. Acting as a cognitive intermediary, the IAI framework enhances human capabilities with AI-driven insights, markedly reducing the complexity and manpower usually needed for such tasks. Furthermore, by automating the problem formulation phase \cite{xi2023rise}, IAI significantly lowers the risk of human error, a crucial aspect of complex network resource allocation. While this innovation offers distinct advantages to new network designers by making it easier to get involved, its main strengths are in making the optimization process more straightforward, efficiently managing network resources, and improving the accuracy and trustworthiness of network models.



\subsection{Proposed Framework}
Accordingly, to effectively generate problem modeling, as shown in Fig.~\ref{fig_PSNC_2}, we propose an IAI-enabled problem formulation framework, which consists of the following units.

{\bf Perception: }The Perception component of the IAI-enabled problem formulation framework is akin to human sensory reception, drawing from diverse sources and modalities. This multi-modal approach enables the IAI to assimilate information from text, visuals, and numerical data. Upon receiving this data, the system employs prompt engineering to transform raw information into structured embeddings. These embeddings are designed to encapsulate the complexity of the input data in a format that is readily interpretable by the IAI agents. Consequently, the Perception component prepares the IAI to comprehend the nuances of the environment and facilitates informed decision-making. In our framework, the Perception component thus serves as the foundational interface between multifaceted data inputs and the IAI's cognitive mechanisms, ensuring that the system's responses are grounded in a comprehensive understanding of the environmental context.

{\bf Brain: }The Brain is the central component of the IAI system, functioning in a three-unit structure: database, storage, and decision-making. The RAG database contains a wealth of searchable academic texts, such as those pertaining to unmanned aerial vehicles (UAVs), reconfigurable intelligent surfaces (RISs), wireless power transfer (WPT), and more, as its training dataset. This corpus of knowledge is transformed into text embeddings that are stored within the IAI's knowledge base.  For the decision-making, the system adopts a plug-in architecture for selecting the appropriate LLM, such as GPT 4, LLAMA 2, Gemini, and Bloom.   When the database unit and the decision-making unit are combined, they form the RAG module. RAG allows the brain to infer, learn, retrieve, and reason, combining its capabilities with database resources to plan and execute actions to provide appropriate strategies for generating problems. For the storage, it is a repository of the agent's historical observations, thoughts, and actions. It enables the agent to effectively recall and apply previous strategies when tackling complex reasoning tasks, similar to how humans draw on memory to navigate unfamiliar situations. The Brain's architecture is integral to the IAI's learning process, providing a contextual understanding and the ability to utilize historical insights for informed decision-making, thereby enhancing the academic rigor of the system's cognitive process.

{\bf Action: }Within the IAI framework, the Action component responds to the Brain's directions and performs tasks in the network environment. This unit of the system follows a clear four-step method. It starts by analyzing network designer inputs, guided by the Brain's insights. Then, it pulls useful information from its knowledge base. Next, the module shapes the network problem into a clear mathematical form. The final step is to present the formulated optimization model. The Action module turns plans into real results, making sure that the IAI's decisions have a direct and practical impact on the network.

{\bf Environment: } The Environment component of the IAI framework is pivotal in encapsulating the application domain for the IAI agent's operation. It represents the real-world network scenarios as described by the network designer, forming the foundation for the IAI's problem formulation process. For instance, in a scenario where a network designer wants to formulate an optimization problem for a RIS-assisted SWIPT network with RSMA, the interaction begins with the network designer's initial request for assistance. The IAI agent then prompts the network designer for specific details, such as the system model, the optimization objective, the decision variables, and any necessary constraints. Through this iterative dialogue, the network designer provides the required information step-by-step. Then, the network designer might describe the network's configuration for optimizing EE, and the IAI agent will use its information to develop an appropriate optimization problem. This interactive process allows the IAI system to thoroughly understand the network designer-defined environment and generate an optimized problem model that aligns with the specific characteristics and requirements of the network scenario. It showcases how the Environment component acts as a collaborative stage, integrating network designer inputs with IAI functionality to create tailored problem-solving approaches for complex network environments.

\begin{figure}[!t]
\centering
\includegraphics[width=0.4\textwidth]{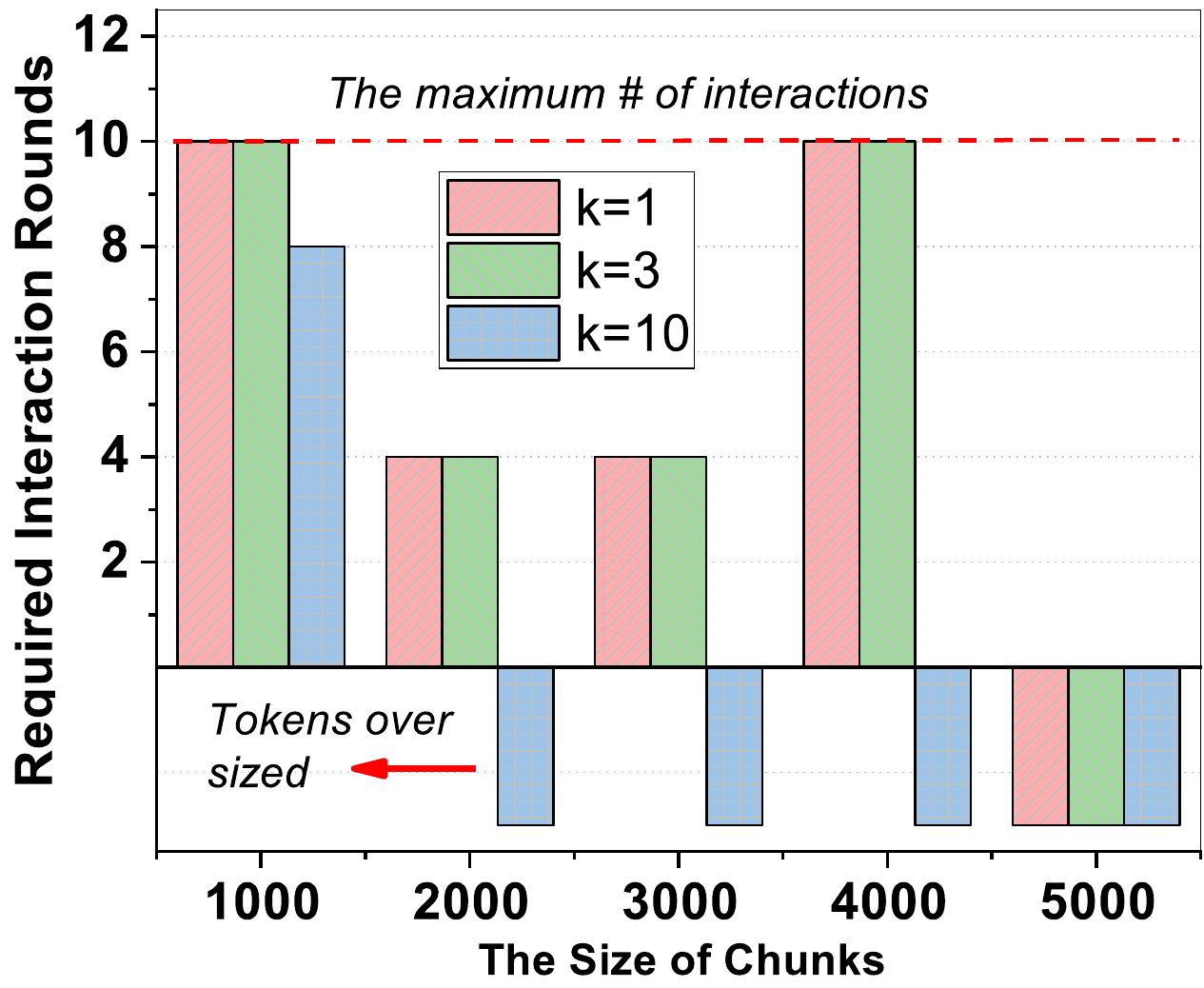}
\caption{The chunk size versus the number of interaction rounds required to solve the network optimization task, where $k$ represents the number of chunks that the LLM puts into context in each round of interaction.}
\label{fig_PSNC_3}
\end{figure}

\begin{figure*}[!t]
\centering
\includegraphics[width=0.75\textwidth, height = 0.75\textwidth]{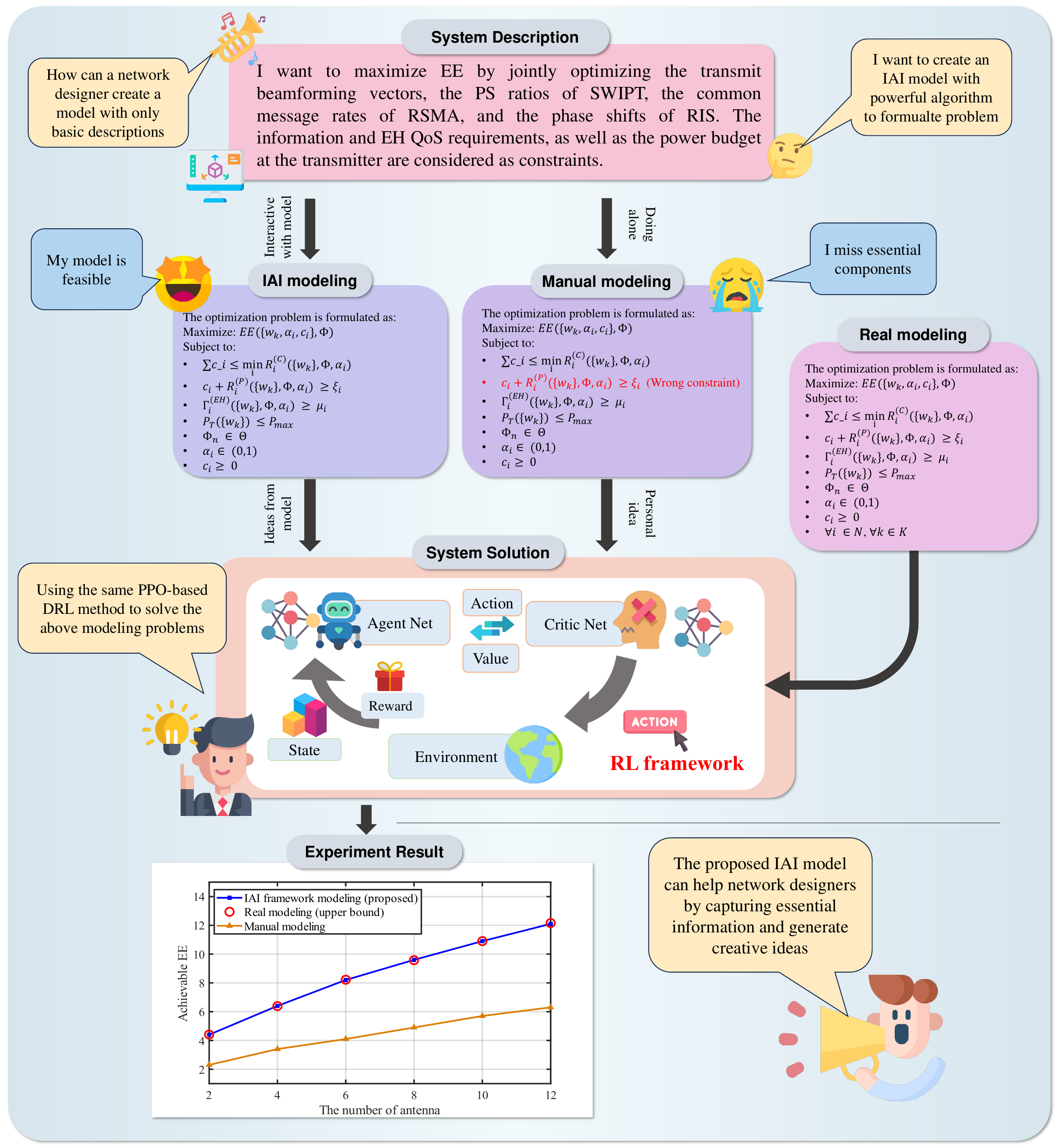}
\caption{Comparison of system performance under various optimization problem generation methods. The figure displays the effectiveness of the proposed IAI framework modeling against traditional real modeling (upper bound) and manual modeling approaches, where the PPO-based DRL method is set as the solution method to demonstrate the performance results.}
\label{fig_PSNC_4}
\end{figure*}
\subsection{Experiments}
\textbf{Experimental Settings}: To explore the effectiveness of the proposed IAI agent, we apply it to solve a real-world network optimization problem.
In our experiments, the pluggable LLM module is implemented by OpenAI APIs for calling the GPT-4 model.
The network-oriented knowledge base and context memory are built on LangChain.

\textbf{Effectiveness of IAI Agent}: As shown in Fig.~\ref{fig_PSNC_2}, the user intends to optimize the energy efficiency in RIS-assisted simultaneous wireless information and power transfer (SWIPT) networks with rate splitting multiple access (RSMA) \cite{zhang2023energy}.
Traditionally, the network designer needs to search numerous articles, select an appropriate model, and apply it to the problem.
With the help of an IAI agent, such a process can be realized automatically through four rounds of user-agent interactions.
To be specific, the user first states the requirements, i.e., formulating a network optimization problem.
Here, the chain-of-thought prompting is applied to enhance the reasoning ability of the IAI agent.
Then, the user further illustrates the specific scenario to the IAI agent, which then returns the step-by-step guidance of the entire optimization process.
Afterward, the users clarify the optimization objective.
Empowered by information retrieval techniques, the IAI agent will jointly search the local knowledge base and perform the inference.
The resulting problem formulation perfectly aligns with the ground truth (i.e., the standard problem formulation from academic journals).
Moreover, the user can acquire detailed explanations of all the factors.

\textbf{Knowledge Base Settings}: With the effectiveness of the IAI agent being verified, we further evaluate the influence of knowledge base settings on the interaction quality.
Specifically, we adjust the chunk size for organizing knowledge embeddings in RAG module and test the number of interaction rounds required for solving the aforementioned task.
As shown in Fig.~\ref{fig_PSNC_3}, when $k$=1 and 3, setting chunk size as 2000 or 3000 can lead to the best interaction quality since the knowledge chunks can be effectively fetched.
In contrast, if the chunk size decreases to 1000, the problem cannot be completely formulated within 10 rounds of interactions because the knowledge that can be referred to is too limited.
Note that extremely large chunk size is also undesirable.
For instance, when the chunk size equals 4000, the problem formulation also fails since the LLM can hardly extract useful knowledge from large chunks.
When the chunk size reaches 5000, the IAI agent cannot even perform inference due to the limited context size.
If increasing $k$ to 10, i.e., allowing the IAI agent to retrieve more chunks in each round, the interaction quality gets improved when chunk size equals 1000.
However, the context oversize error happens in all other settings.

\textbf{Performance Comparison}:  To demonstrate the efficacy of our proposed IAI framework, Fig.~\ref{fig_PSNC_4} shows an example to present a comparison of different ways for generating optimization problems. It is evident that the first step in creating an optimization problem involves providing a detailed problem description. Our IAI framework then interacts with this description, as illustrated in Fig.~\ref{fig_PSNC_2}, to automatically derive the corresponding optimization problem. In contrast, network designers often rely on their experience to manually create optimization problems. However, this manual approach can be prone to errors, especially when designers are novices or unfamiliar with certain aspects, such as overlooking or incorrectly defining constraints (e.g., decoding RSMA constraints in our example). Such oversights can lead to flawed optimization problems. Upon applying the same PPO-based optimization method for solving, it is observed that the performance of algorithms under our framework closely matches that of the original real problem formulation (i.e., in \cite{zhang2023energy}), whereas manually designed optimization problems yield the lowest results. This difference is attributed to the strength of the IAI framework, which leverages LLM capabilities and features within LangChain to generate precise optimization problems through an interactive process. Therefore, the IAI framework ensures a more accurate and efficient problem formulation compared to traditional manual methods.

\textit{Lesson learned:}  Through the above experiments, we can observe that the IAI goes beyond GAI, which only considers the quality of generated content.
Instead, the IAI agent is designed to help users accomplish specific tasks.
Therefore, not only the core LLM but also the settings for other modules that guarantee the smoothness and quality of interaction (e.g., the knowledge base and memory) should be well crafted.

\section{Future Directions}
In this section, we outline three main future directions for the improvement of IAI-enabled networking. 

{\bf Integration with Emerging Technologies:}
IAI can enhance B5G/6G networks, particularly in real-time data processing, adapting to bandwidth and latency needs. Edge computing boosts IoT and smart city operations. Blockchain integration uses IAI for more secure, decentralized networks. IAI also makes semantic communications more context-sensitive and improves ISAC systems' accuracy. These enhancements show IAI's pivotal role in modern network technologies.

{\bf Security Aspects of IAI-Enabled Networks:}
Improving security in IAI networks is crucial. Future research should focus on IAI-driven protocols for early threat detection and adaptive response, automating security management, and evolving with emerging threats.

{\bf Evaluation of IAI systems:}
It is important to design a model that can evaluate IAI. In the future, AI evaluation criteria should be given to evaluate models generated by IAI instead of human user evaluation.

\section{Conclusion}
In this article, we have explored the integration and enhancement of IAI in networking. We have proposed a problem formulation framework by using IAI with RAG, where the effectiveness of the framework was verified through simulation results. Finally, some potential research directions for IAI-based networks were outlined.

\bibliographystyle{IEEEtran}
\bibliography{mylib}

\end{document}